# Valley splitting by extended zone effective mass approximation incorporating strain in silicon

Jinichiro Noborisaka, Toshiaki Hayashi, Akira Fujiwara, and Katsuhiko Nishiguchi

*NTT Basic Research Laboratories, NTT Corporation,*

*3-1 Morinosato-Wakamiya, Atsugi, Kanagawa 243-0198 Japan*

Abstract

Silicon metal-oxide-semiconductor field effect transistors (MOSFETs) fabricated on a SIMOX (001) substrate, which is a kind of silicon on insulator (SOI) substrate, that is annealed at high temperature for a long time are known to exhibit large valley splitting, but the origin of this splitting has long been unknown. Extended zone effective-mass approximation (EMA) predicts that strain significantly affects valley splitting. In this study, we analyzed valley splitting based on this theory and found that the shear strain along <110> of approximately 5% near the buried oxide (BOX) interface is a promising source for large valley splitting.

The valley degrees of freedom have long been studied as an important factor that affects electron scattering and relaxation in silicon. This in particular is considered as a critical parameter that must be precisely controlled in actualizing spin-based quantum computers [1, 2]. It has long been known that valley splitting occurs at the normal MOS interface in silicon [3-5], and its magnitude is typically in the order of several hundred microelectron-volts under typical operating-field strengths of silicon MOSFETs [6-8]. In contrast, MOSFETs fabricated on specially treated silicon substrates exhibit huge valley splitting in the order of tens of millielectron-volts [9, 10]. Silicon is a widely used material in electronic devices, but the origin of valley splitting, one of the fundamental physical phenomena of silicon, is still controversial.

In bulk silicon, electrons in the conduction band are six-fold degenerate because the conduction band minima do not lie on the Γ point but on the orthogonal [100], [010], and [001] axes as shown in Fig. 1(a). If we introduce <001> spatial confinement to form a so-called quantum well (QW), the electrons split into two groups of valley states. When the confinement is not so strong, the valleys with lighter EM along <001> become 4-fold degenerate, and the remaining valleys with heavier EM are doubly degenerate. This doubly-degenerate state is the ground state of the QW [11]. By applying an external field or introducing a physical structure to establish even stronger confinement along <001>, the doubly-degenerate state exhibits further splitting. This energy splitting is referred to as valley splitting. Valley splitting is fundamentally attributed to the allowance of valley degeneracy due to the difference in wavenumber k. When a very strong confinement is introduced, the wave functions belonging to states $-k_0$ and $+k_0$ superimpose a wide range of wavenumber states in the momentum space. The probability densities of electrons in each valley eventually overlap in the momentum space as shown in Fig. 1(b). When an overlap of valley states occurs, the respective states interfere, hybridize, and exhibit energy splitting.

Research on valley splitting has a long history. In the integer quantum Hall effect in silicon, a splitting caused by degrees of freedom distinct from spin was observed [3-5]. This splitting was found to be associated with the valley degrees of freedom. At this time, basic theories on valley splitting based on EMA were developed, and valley splitting at the normal Si/SiO$_2$ interface was well explained by these theories [6-8,12-15]. In the 1990s, papers reporting valley splitting several tens of times larger than that at the normal Si/SiO$_2$ interface [9, 10] attracted much attention, where MOSFETs were fabricated on SIMOX (001) substrates annealed at high temperatures for a long time. Subsequently, large valley splitting was verified based on various theories such as tight binding models [16, 17], EMA incorporating a new perturbation term fitted by a tight binding theory so as to avoid the problems in describing microscopic phenomena [18], and EMA combined with first-principle calculations [19]. There were also theories incorporating interface states [20, 21], abrupt interfaces [22], and superlattices [23, 24]. The effect of substrate tilting [14] was also reconsidered. However, none of these theories have provided a clear answer to the origin of large valley splitting.

In this paper, we analyze experimental results on large valley splitting based on the theoretical framework of the extended zone EMA reported in Ref. 6. The paper incorporates the effects of strain that have been overlooked in many previous papers demonstrating that the introduction of shear strain significantly enhances valley splitting caused at the X point. Based on this theory, it is predicted that a compressive strain of approximately 5% along <110> is introduced near the BOX and SOI interface of the SIMOX substrate.

Figure 1(c) shows the device structure. This device uses a SIMOX (001) substrate [25] as the starting material. First, the SIMOX substrate is annealed at 1350°C for 40 hours in an atmosphere of 0.5% $O_2$ in Ar to improve the flatness of the BOX/SOI interface [26, 27]. The surface oxide film formed by 40-hour annealing is then removed using hydrofluoric acid, and after adjusting the thickness of the SOI ($t_{SOI}$), the front gate oxide (FOX) film is formed by thermal oxidation at 900°C. Next, by depositing gates on both sides of the FOX and the substrate, we fabricate double-gate structure MOSFETs. Using the front gate (FG) and back gate (BG), we adjust the spatial distribution of the electron-wave function inside a thin SOI, *i.e.*, a QW, and observe the valley splitting at the FOX or BOX interface. The thicknesses of the FOX ($t_{FOX}$) and BOX ($t_{BOX}$) are 20 nm and 400 nm, respectively, and that of the QWs ($t_{SOI}$) are 6 nm and 9 nm. The gate length ($L_g$) and width ($W_g$) are 400 μm and 200 μm, respectively. Note that the BOX and FOX formation methods are remarkably different because the formation of the BOX is performed around unique temperatures exceeding the glass transition of $SiO_2$ or temperatures near the melting point of silicon.

Figure 2(a) shows typical drain current $I_D$ versus FG voltage $V_{FG}$ characteristics of the fabricated MOSFETs ($t_{SOI}$ = 9 nm). All of the following measurements were conducted at 4.5 K. For BG voltage ($V_{BG}$) > 0 and $V_{FG}$ < 0, only the channel at the BOX interface is open, where a distinct kink structure is observed (indicated by arrows). This structure reflects the two-dimensional density of states (2D-DOS) in the valley-coupled condition. This structure is related to the valley splitting that we focus on in this paper. A negative differential structure appears at $V_{FG}$ ~ 0 V, indicating a newly opened channel at the FOX interface. In contrast, for $V_{BG}$ < 0, only the FOX interface channel is open and no particular structure appears. Figure 2(b) shows a contour plot of $\partial^2 I_D / \partial V_{FG}^2$. For $V_{BG}$ < 0, a single-line structure due to the threshold of the FOX interface channel is observed, while for $V_{BG}$ > 0 a distinct splitting structure appears. This structure originates from the $I_D$-$V_{FG}$ kink described above, which corresponds to valley splitting. The lower line of $\partial^2 I_D / \partial V_{FG}^2$ represents the onset of the electron occupation of the 2D state originating from the bonding orbital of the valley-coupled state, and the upper line represents its antibonding orbital.

Figure 2(c) shows the $V_{BG}$ dependence of the valley splitting for each $t_{SOI}$ thickness. The sizes of valley splitting are obtained by 2D-DOS of silicon and electron densities, which are given by the capacitance of FG ($C_{FG}$= $\varepsilon_{SiO2}$ $\varepsilon_0$ / ($t_{FOX}$ + $t_{SOI}$ $\varepsilon_{SiO2}$ /$\varepsilon_{Si}$)) and the $V_{FG}$ difference (Δ$V_{FG}$) in $\partial^2 I_D / \partial V_{FG}^2$, where $\varepsilon_{SiO2}$, $\varepsilon_{Si}$, and $\varepsilon_0$ represent the relative permittivity of $SiO_2$, silicon, and the permittivity in a vacuum, respectively. This figure also shows the electric field of the SOI ($F_{SOI}$) corresponding to each $V_{BG}$. This $V_{BG}$ to $F_{SOI}$ conversion is obtained by comparing the $V_{BG}$ dependence of the transverse optical phonon peak given by electroluminescence with the numerical simulation of the quantum confined Stark shift [28], where electroluminescence is observed using the pn contacts of the same device. In the device with $t_{SOI}$ = 6 nm shown in Fig. 2(c), valley splitting is indeed approximately 30 times larger than that expected at the normal Si/SiO$_2$ interface. For $V_{BG}$ < 0, electrons are squeezed at the normal Si/SiO$_2$ interface (FOX), resulting in an energy splitting of a few hundred microelectron-volts the value of which is so small that we were unable to observe a distinct structure. On the other hand, when $V_{BG}$ > 0, electrons are squeezed at the BOX/SOI interface, and in this state, a huge valley splitting occurs that cannot be explained by conventional theory. Figure 2(c) shows that the difference in valley splitting due to the difference in $t_{SOI}$ is small. This is because, even theoretically, the valley splitting dependence on thickness appears only near a zero electric field, and the thicker the film, the lower the film-thickness dependence. Therefore, the difference in film thickness is almost negligible on the scale we are measuring.

According to the theory by Cardona *et al.*, the states at any wavenumber of the conduction band (CB) in silicon can be expanded by three gamma bases: $\Gamma_{15}$, $\Gamma_1^u$, and $\Gamma_1^l$ [29]. In the extended zone EMA by Ohkawa *et al.*, the state of $\Gamma_1^l$ is further perturbatively treated, and the lowest CB is expanded only with the two states of $\Gamma_{15}$ and $\Gamma_1^u$. According to this theory, valley splitting is expressed by

$$\Delta E = \sqrt{(\Delta E_\Gamma)^2 + (\Delta E_X)^2} \qquad (1)$$

where $\Delta E_\Gamma$ and $\Delta E_X$ represent valley splitting originating from the valley coupling via the $\Gamma$ and X points, respectively. Here, $\Delta E_\Gamma$ is expressed by

$$\Delta E_\Gamma = \varepsilon_\Gamma S_\Gamma \qquad (2)$$

$$S_\Gamma = \left| \int e^{i2k_0 z} \xi(z)^2 dz \right| \qquad (3)$$

where $\varepsilon_\Gamma$ represents the energy difference between $\Gamma_1^u$, and $\Gamma_{15}$ as shown in Fig. 3(a). Here, $S_\Gamma$ represents the overlap integral of envelope function $\xi(z)$ corresponding to the $+k_0$ and $-k_0$ valley states in the QW, which can be interpreted as the Fourier transform of the square of the envelope function (Fig. 1(b)). Term $k_0$ is the wavevector that gives the valley minimum point, $k_0 = 0.85<001>2\pi/a_0$, where $a_0$ is the lattice constant of silicon. Therefore, valley splitting resulting from the $\Gamma$ point is proportional to the $2k_0$ component of the Fourier coefficients of the envelope function squared multiplied by $\varepsilon_\Gamma$.

In contrast, the valley splitting caused by the X point is expressed by the following equation, assuming that in-plane momentum $k_{//}$ is sufficiently small.

$$\Delta E_X = 2\Xi'_u e_{xy} S_X \qquad (4)$$

$$S_X = \left| \int e^{i(2k_0 - 4\pi/a_0)z} \xi(z)^2 dz \right| \qquad (5)$$

where $\Xi'_u$ is the shear-strain deformation potential, and $e_{xy}$ represents the shear strain along <110>. In silicon, the CB is degenerate at the X point due to glide symmetry, and this symmetry is broken when shear strain is applied as shown in Fig. 3(a). Note that hydrostatic strain has no effect on the valley splitting via the X point because it only shifts the $\Delta_1$ and $\Delta_2$ bands in the same direction [30]. In the extended zone scheme, the neighboring energy minima of the 2nd Brillouin zone (BZ) across the X point corresponds to the valley of $-k_0$ as shown in Fig. 3(a).

Comparing $\Delta E_\Gamma$ and $\Delta E_X$ in Eqs. (2) and (4), respectively, both equations contain overlap integral S between the envelope functions corresponding to $-k_0$ and $+k_0$. The magnitude of this overlap integral is inversely proportional to the distance in the wavenumber-space ($\Delta k$) of these valleys. The distance at the $\Gamma$ point is $2k_0 = (2\pi/a_0) \times 17/10$, while at the X point it is $|2k_0 - 4\pi/a_0| = (2\pi/a_0) \times 3/10$, which is approximately six times closer to the X point. Consequently, $S_X$ is approximately two orders of magnitude larger than $S_\Gamma$, and the valley splitting via the X point becomes significant when the strain along <110> ($e_{xy}$) is nonzero.

Details of the calculation method for valley splitting are given in the supplementary materials. In short, valley splitting is calculated using Eqs. (2) to (5) and real-space envelope function $\xi(z)$ obtained by numerical solution of the single-band EM equation using the finite element method.

Figures 3(b) and 3(c) show the calculated electric field dependence of $\Delta E_\Gamma$ and $\Delta E_X$ at each $t_{SOI}$, respectively. In $\Delta E_X$, shear strain $e_{xy}$ is set to 5%. Both graphs show that valley splitting increases as $t_{SOI}$ decreases or $F_{SOI}$ is increased. This is because under both conditions, the $2k_0$ component of the overlap integral on k-space takes a large value due to the stronger confinement of the electrons in the QW. When focusing on $t_{SOI}$ dependence near the zero electric field, the large $t_{SOI}$ exhibits linearity in terms of $F_{SOI}$ dependence, while the extremely small $t_{SOI}$ exhibits nonlinearity. This is because when $t_{SOI}$ is small, the barrier potential becomes the main cause of electron confinement, but as $t_{SOI}$ increases, the triangular potential caused by the electric field becomes the primary factor for electron confinement. Therefore, when $t_{SOI}$ is small, nonlinearity is confirmed because the effective well width changes as it transitions from well confinement to electric-field confinement. Additionally, since $\Delta k$ is much closer at the X point, nonlinearity near the zero electric field is emphasized even in QWs with a relatively large $t_{SOI}$ in $\Delta E_X$. Also, comparing $\Delta E_\Gamma$ and $\Delta E_X$ in Figs. 3(a) and 3(b), $\Delta E_X$ is much larger than $\Delta E_\Gamma$. This is because the shear strain is set to a large value and the $\Delta k$ of each valley at the $\Gamma$ and X points described above, where $\Delta k$ is much closer to the X point.

Figure 3(d) shows the $F_{SOI}$ dependence on each shear strain of the overall valley splitting, which is the sum of the valley splitting due to the $\Gamma$ and X points expressed in Eq. (1). For a near zero electric field, valley splitting is strongly affected by the valley splitting through the X point and it exhibits a nonlinear change with respect to the electric field, but basically the size of the valley splitting is simply proportional to the shear strain.

Figure 4 shows valley splitting overlaid on a contour plot with shear strain chosen to fit the experiment. Although nonlinearities are noticeable in the theoretical value near the zero electric field, the experimental and theoretical values for 6 nm and 9 nm are in relatively good agreement. The reason why the nonlinearity is not confirmed in the experiment may be that the centroid of the wave function is far from the BOX interface near the zero electric field, so the strain effect ($\Delta E_X$) is weak. If a similar fitting is performed with valley splitting through point $\Gamma$ only, the electric field must exceed the breakdown of $SiO_2$, which would not be consistent with the experiment. Figure 4 shows that introducing a compressive strain along <110> of 4.5% at $t_{SOI}$ = 6 nm and 5% at $t_{SOI}$ = 9 nm as the shear strain agrees well with the experiment results, indicating that shear strain along <110> near the BOX interface is promising as the origin of large valley splitting. This strain value is very large from a general strain viewpoint, but this would not cause silicon to fracture.

Here, we discuss the validity of considering strain as the origin of the large valley splitting. The results obtained by Niida *et al.* [31], who measured the mobility of MOSFETs fabricated on SIMOX substrates subjected to high-temperature annealing similar to the conditions presented herein, revealed the following characteristics when squeezing electrons towards at the BOX interface. The electron mobility decreases compared to the case of squeezing at the FOX interface with the same electron density, while the hole mobility increases. This is consistent with the behavior of the <110> electron [32] and hole mobility [33] when introducing <110> compressive shear strain in (001) silicon. Since the mobility can vary due to various factors such as interface quality and impurity concentration, it is difficult to compare mobility by itself. However, the agreement between the Ohkawa theory and experiments, and the consistency of the changes in both electron and hole mobility when <110> directional compressive strain is applied suggest that the <110> directional compressive strain near the BOX interface is a promising candidate as the origin of large valley splitting. The origin of the massive strain near the BOX interface remains unclear, and further detailed investigation is needed to determine whether strain actually exists and, if so, its physical origin. Finally, we note that high reproducibility in the size of large valley splitting has been obtained among the multiple devices measured so far even among different wafers.

We demonstrated that large valley splitting observed in SIMOX substrates subjected to high-temperature long-duration annealing can be explained by the extended zone EMA proposed by Ohkawa *et al.* In the theory, valley splitting via the X point may play an important role and have a significant impact on transport phenomena in the presence of shear strain in the inversion layer of MOSFETs. Further analysis based on this theory shows that the experimental large valley splitting is well reproduced when a shear compressive strain of approximately 4-5% along <110> is introduced into the SOI layer at the BOX interface.

## Acknowledgements
We thank Professor Y. Tokura and Professor Y. Kageshima for their helpful discussions.

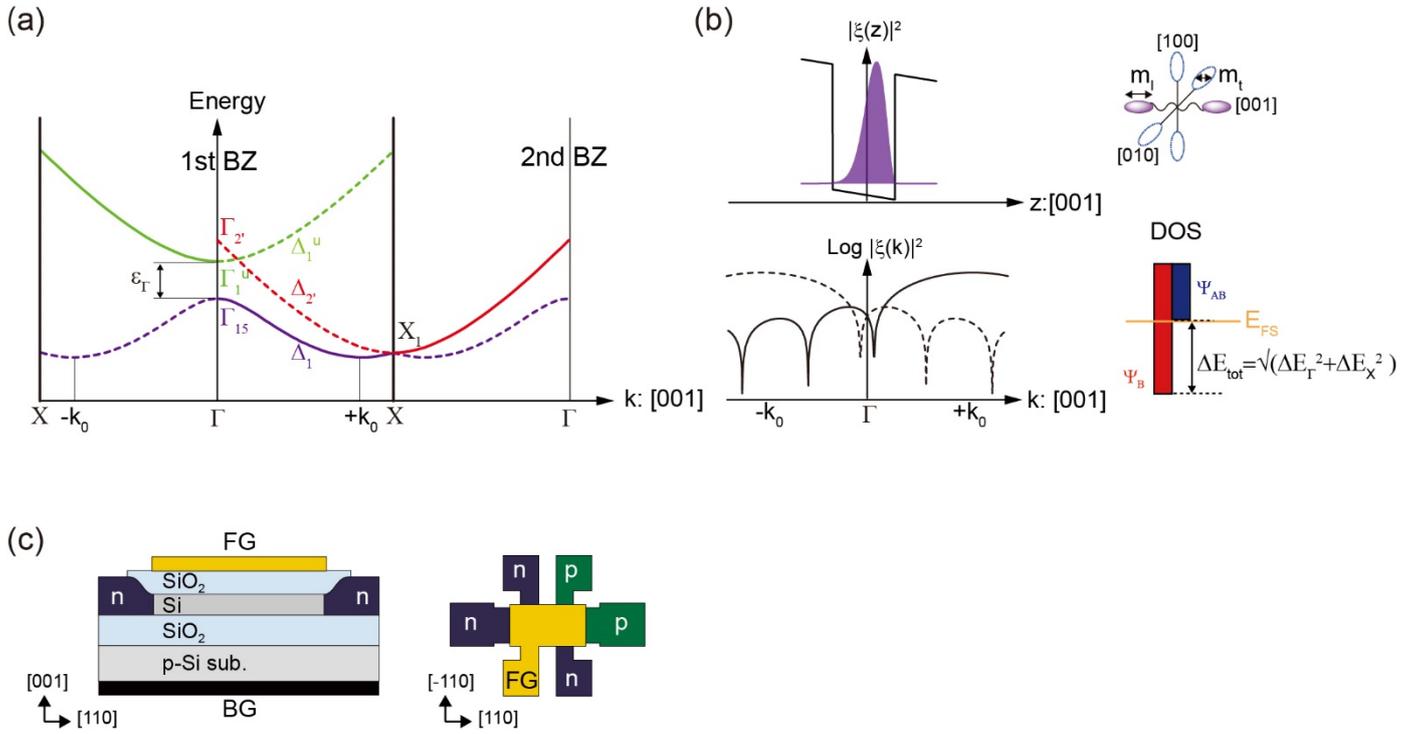

Fig. 1 (Color online) (a) (Top) Band structure of silicon: Solid line is $+k_0$ and dotted line is the contribution by $-k_0$ valley. The band structure at the $\Gamma$ point in actual silicon is different from that in this figure due to the anti-intersection of $\Gamma_1^u$, $\Gamma_2'$. (b) (Top left) Real space representation of the envelope function squared. (Bottom left) Wavenumber-space representation of the envelope function squared (Solid and dotted lines are from $+k_0$ and $-k_0$ valleys, respectively). (Top right) Equi-energy surface of CB. For (001) quantum wells, doubly degenerate valleys with heavy EM $m_l$ along <001> form the ground level. (Bottom right) Schematic diagram of valley splitting, where $\Psi_B$ and $\Psi_{AB}$ represent bonding and antibonding states, respectively. The valley splitting observed in the experiment, $\Delta E_{tot}$, is the sum of squares of valley splitting $\Delta E_\Gamma$ by $\Gamma$ point and $\Delta E_X$ by X point. (c) Device structures: (Left) Section view. (Right) Overview

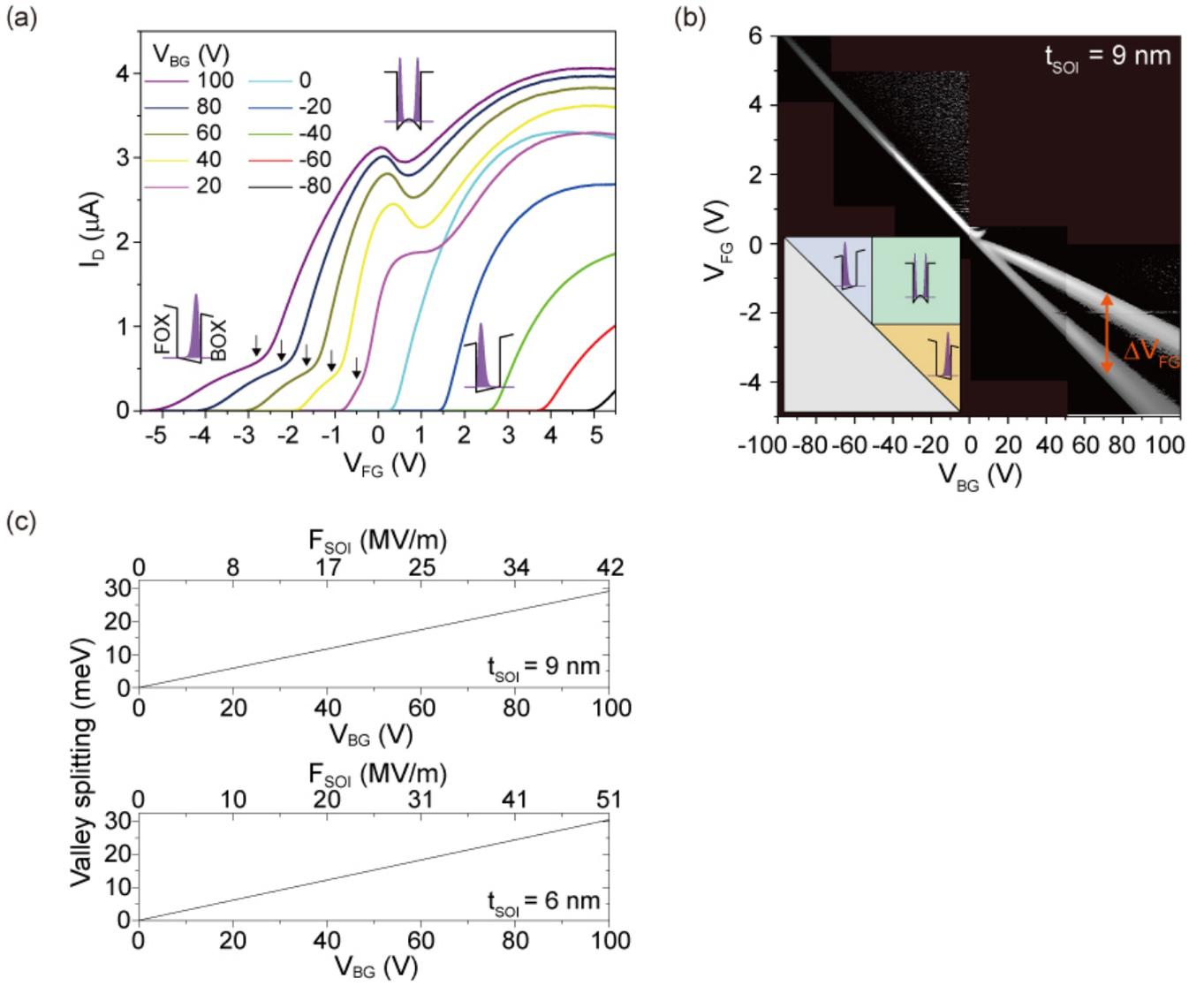

Fig. 2 (Color online) (a) $I_D$-$V_{FG}$ characteristics with $t_{SOI}$ = 9 nm: Kink structures associated with valley splitting are identified in $V_{BG} > 0$ and $V_{FG} < 0$ (indicated by arrows). (b) $\partial^2 I_D / \partial V_{FG}^2$ with $t_{SOI}$ = 9 nm: $V_{BG}<0$ confirms a single structure with FOX side channel threshold. For $V_{BG} > 0$, a clear splitting structure is seen. (Inset) Electron distribution in QW in each region. (c) Measured valley splitting with $t_{SOI}$ = 6 and 9 nm.

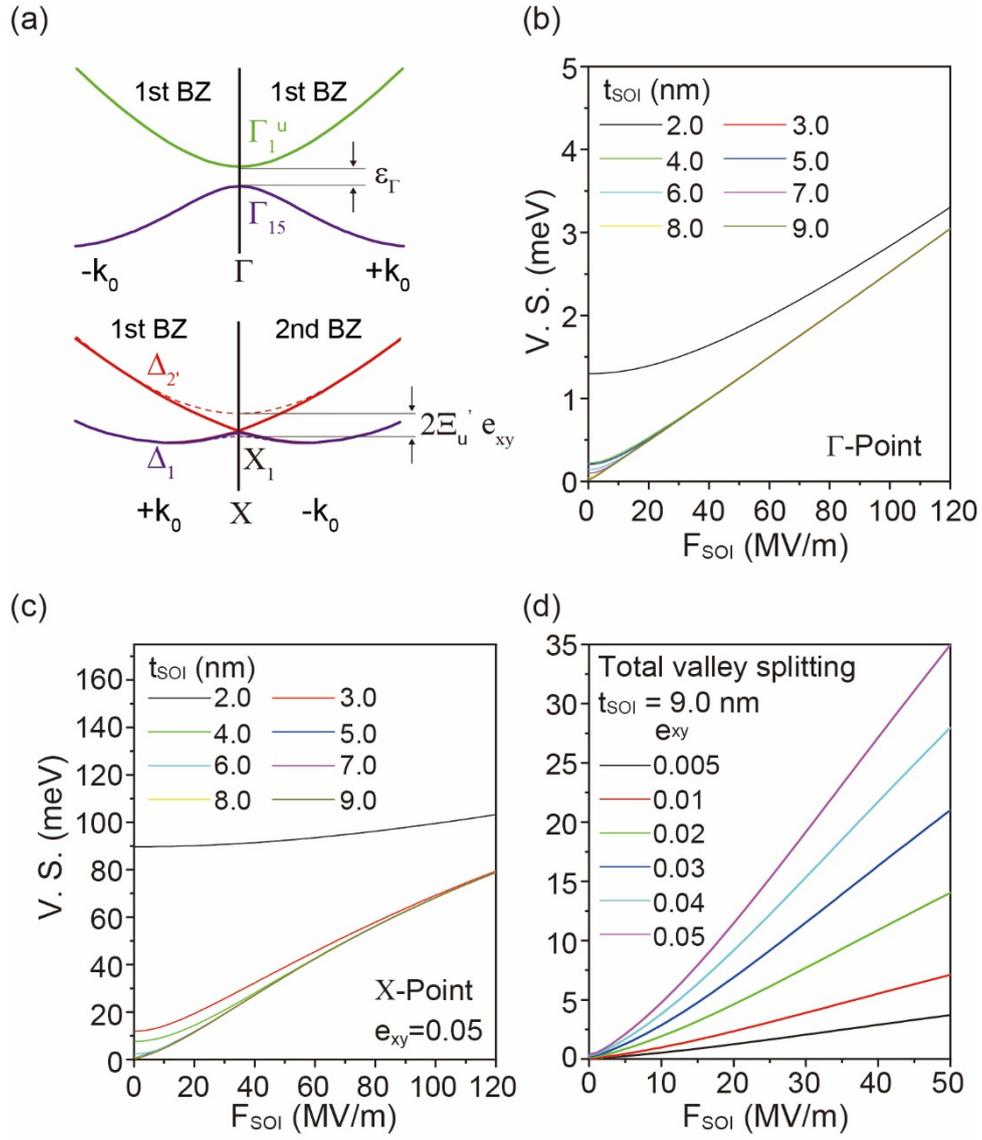

Fig. 3 (Color online) (a) Energy splitting at Γ and X points (dotted line: with shear strain): When shear strain or in-plane momentum $k_{//}$ is small, $\Delta_1$ and $\Delta_{2'}$ bands degenerate at X point due to glide symmetry. (b) Valley splitting (V. S.) $\Delta E_\Gamma$ due to Γ point. (c) Valley splitting $\Delta E_X$ due to X point (shear strain is 5%). (d) Whole valley splitting taking the sum of squares of $\Delta E_\Gamma$ and $\Delta E_X$.

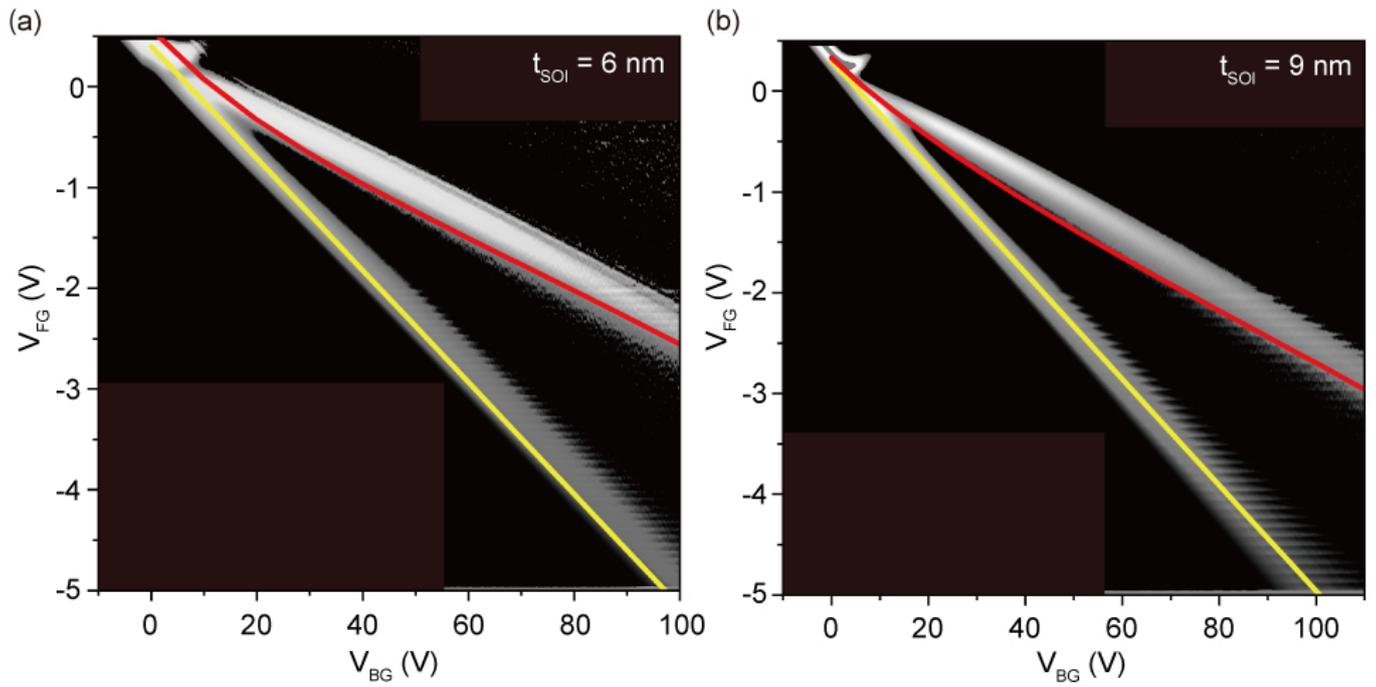

Fig. 4 (Color online) (a) Valley splitting with $t_{SOI}$ = 6 nm: (Comparison between experiment and theory) Solid line (theory) is for shear strain exy = 0.045. (b) Valley splitting with $t_{SOI}$ = 9 nm: Solid line (theory) is for shear strain exy = 0.05.

# Supplemental Material to Valley splitting by extended zone effective mass approximation incorporating strain in silicon

Jinichiro Noborisaka, Toshiaki Hayashi, Akira Fujiwara, and Katsuhiko Nishiguchi

*NTT Basic Research Laboratories, NTT Corporation,*

*3-1 Morinosato-Wakamiya, Atsugi, Kanagawa 243-0198 Japan*

The calculation method for valley splitting is described below. Ohkawa *et al.* [7] provides a solution for valley splitting by solving a two-band effective mass (EM) equation, *i.e.*, real-space envelope function $\xi(z)$ in Eqs. (3) and (5) that is obtained by solving the two-band EM equations below simultaneously.

$$[H_\Gamma + V(z)]\xi(z) = \epsilon\xi(z)$$

$$H_\Gamma = \begin{pmatrix} \frac{\epsilon_\Gamma}{2} - \frac{\partial^2}{\partial z^2} & -iT\frac{\partial}{\partial z} \\ -iT\frac{\partial}{\partial z} & -\frac{\epsilon_\Gamma}{2} - (1+S)\frac{\partial^2}{\partial z^2} \end{pmatrix} \quad (S1)$$

In Ref. 7, two series of eigenfunctions, $\xi_{15}(z)$ and $\xi_1^u(z)$, are obtained using the variational method. However, this solution is laborious to solve. Therefore, in this paper, valley splitting is calculated using the envelope function obtained from the numerical solution of the single-band EM equation. We examine the validity of this treatment.

In an infinite quantum well (QW) without an electric field, an analytical solution for the two-band EM equation for valley splitting exists [11], where the valley splitting is given by

$$\Delta E_\Gamma = \epsilon_\Gamma \pi^2 |\sin(k_0 t_{SOI})| / t_{SOI}^3 k_0^3 \quad (S2)$$

Figure S1 shows the analytical solution for the $t_{SOI}$ dependence on valley splitting in the infinite QWs expressed by Eq. (S2) and $\Delta E_\Gamma$, which is obtained using Eq. (2) and real-space envelope function $\xi(z)$. Here, $\xi(z)$ is calculated based on the numerical solution of the following single-band EM equation using the finite element method with the EM of silicon along [001] as $m_l$, its $SiO_2$ as $m_{SiO2}$, and the barrier height as $V_B$. Table S1 gives the parameters used in the numerical calculation. Here, $V_B$ is set to 10 eV to reproduce the infinite QW.

$$\left[-\frac{\hbar^2}{2m^*}\frac{\partial^2}{\partial z^2} + V(z)\right]\xi(z) = \epsilon\xi(z) \quad (S3)$$

where $m^* = m_l$ or $m_{SiO2}$ and $V(z) = 0$ or $V_B$. Figure S1 shows that the numerical calculation obtained from the single-band agrees very well with the two-band analytical solution. This indicates that the effect of the higher band, $\Gamma_1^u$, on the shape of the envelope function is fairly well considered by $m_l = 0.98m_0$ in the single-band EM equation. Since barrier potential $V(z)$ does not directly couple $\Gamma_{15}$ and $\Gamma_1^u$ in the EMA framework, it is reasonable to incorporate the influence of the higher band in the EM even in a finite QW. Therefore, in this paper, valley splitting is computed from real-space envelope function $\xi(z)$ obtained by numerical solution of single-band EM equation (S3) and Eqs. (2) to (5). Figure S1 also shows that the valley splitting oscillates with respect to $t_{SOI}$. This is an interference effect observed between the well width ($t_{SOI}$) and the wavelength given by the inverse of $k_0$.

Since the valley splitting at the X point is obtained from the overlap integral $S_x$ via the X point and shear strain $e_{xy}$, it is calculated using $\xi(z)$ obtained by solving Eq. (S3) in the same way as the $\Gamma$ point.

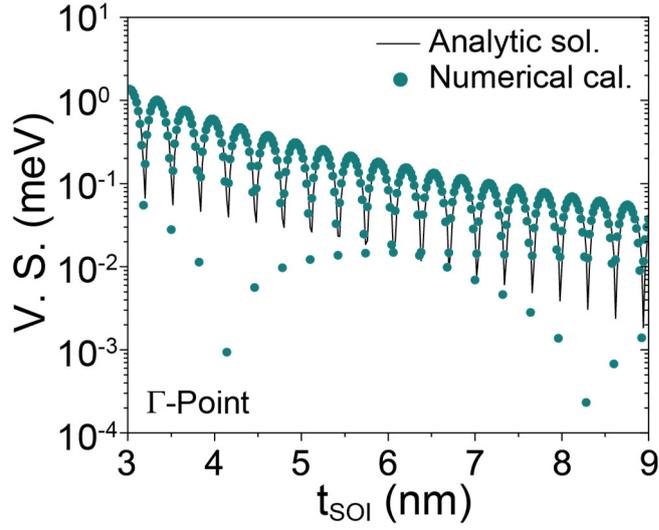

Fig. S1 Valley splitting in an infinite QW without electric field:
The analytical solution of the $t_{SOI}$ dependence on valley splitting is given by Eq. (S2), while the numerical results are given by Eq. (2) and $\xi(z)$ obtained by solving a single-band EM equation.

| Table I. Constants used in this paper |
|---|
| $m_l = 0.98\, m_0$ |
| $m_{SiO2} = 0.98\, m_0$ |
| $V_B$ (finite / infinite) = 3.1 / 10.0 eV |
| $\varepsilon_\Gamma = 3.65$ eV |
| $\Xi_u' = 5.7$ eV |
| $a_0 = 5.43$ Å |